%% file: egpaper_final.tex
\documentclass[10pt,twocolumn,letterpaper]{article}

\usepackage{iccv}
\usepackage{times}
\usepackage{epsfig}
\usepackage{graphicx}
\usepackage{amsmath}
\usepackage{amssymb}
\usepackage{xcolor}
\usepackage{multirow}
\usepackage{adjustbox}
\usepackage{float}
\usepackage{pifont}
\usepackage{svg}
\usepackage{algorithm}
\usepackage{algpseudocode}
\usepackage{tikz}
\usepackage{booktabs}
\usepackage{subcaption}
\usepackage[breaklinks=true,bookmarks=false]{hyperref}

\iccvfinalcopy 

\def\iccvPaperID{21} 

\ificcvfinal\fi
\definecolor{myGreen}{RGB}{34, 139, 34}
\definecolor{Red}{RGB}{204, 0, 0}
\definecolor{Blue}{RGB}{0, 0, 204}
\begin{document}

\title{Gen4D: Synthesizing Humans and Scenes in the Wild}
\newcommand{\proposed}{\texttt{\textbf{Gen4D}}}

\author{\parbox{16cm}{\centering
    {\large Jerrin Bright$^{1,3}$, Zhibo Wang$^{1,3}$, Yuhao Chen$^{1,3}$, Sirisha Rambhatla$^{2,3}$,\\ John Zelek$^{1,3}$, David Clausi$^{1,3}$ }\\
    {\normalsize
    $^1$ Vision and Image Processing Lab\\
    $^2$ Critical ML Lab\\
    $^3$ University of Waterloo, Canada\\
    {\tt\small {\{jerrin.bright, zhibo.wang, yuhao.chen1, sirisha.rambhatla, jzelek\}}@uwaterloo.ca}
    }
}}
\ificcvfinal\thispagestyle{empty}\fi

\twocolumn[{%
\renewcommand\twocolumn[1][]{#1}%
\vspace{-1em}
\maketitle
    \begin{center}
    \centering 
    \includegraphics[width=\linewidth]{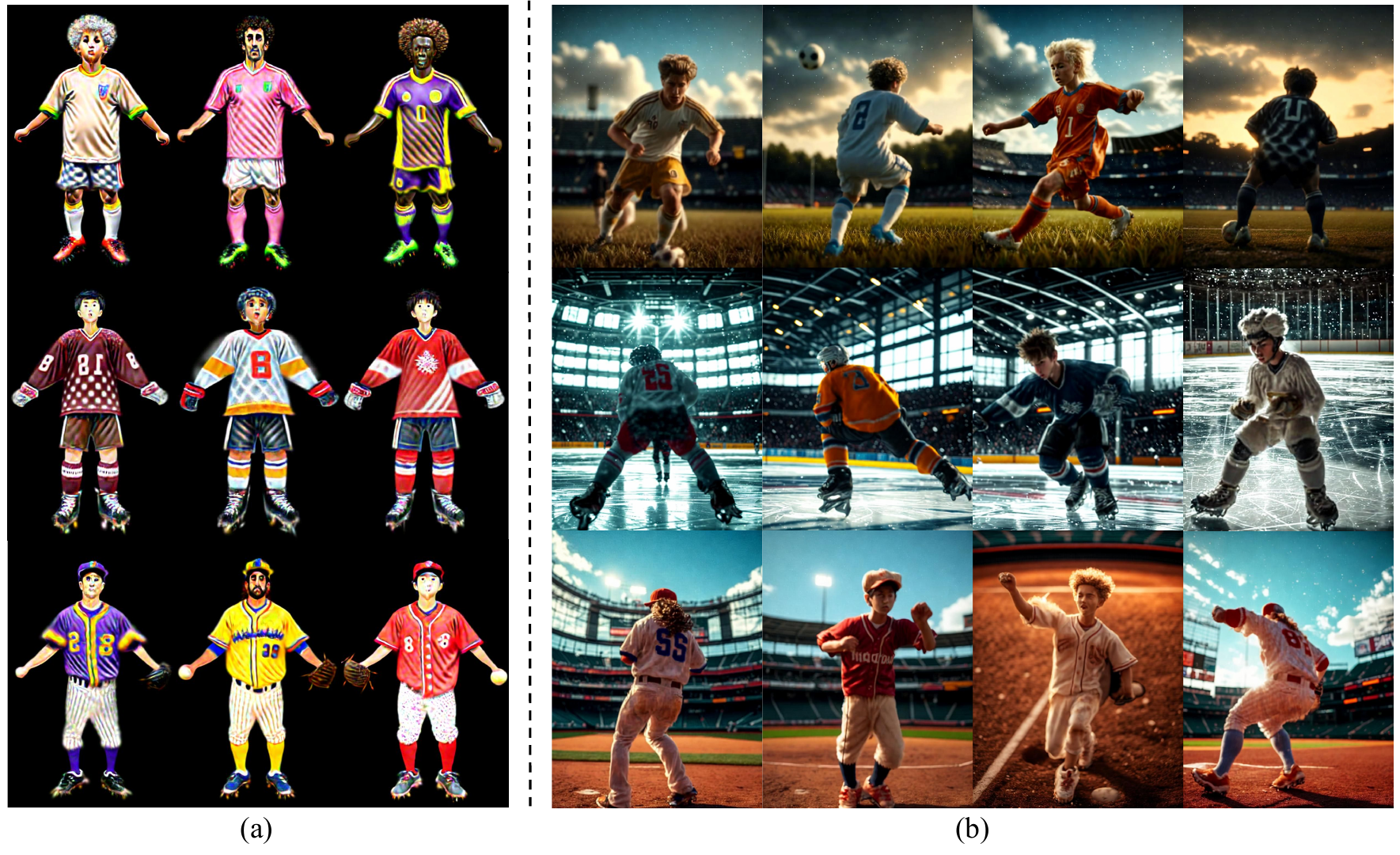}
    \vspace{-25px}
    \captionof{figure}{\textbf{Overview of the {SportPAL} dataset generated using our proposed framework, \textbf{\proposed}.} 
    \textbf{(a)} Examples of diverse canonical avatars synthesized via diffusion-guided prompt modeling, illustrating variation in clothing, body shape, and appearance. 
    \textbf{(b)} Final synthetic frames rasterized with motion-driven avatar animation and human pose-aware backgrounds, demonstrating realistic lighting, shadows, and foot-ground interaction.
    }\label{fig:teaser}
    \end{center}
}]

\begin{abstract}
    Lack of input data for in-the-wild activities often results in low performance across various computer vision tasks. This challenge is particularly pronounced in uncommon human-centric domains like sports, where real-world data collection is complex and impractical. While synthetic datasets offer a promising alternative, existing approaches typically suffer from limited diversity in human appearance, motion, and scene composition due to their reliance on rigid asset libraries and hand-crafted rendering pipelines. To address this, we introduce {\proposed}, a fully automated pipeline for generating diverse and photorealistic 4D human animations. {\proposed} integrates expert-driven motion encoding, prompt-guided avatar generation using diffusion-based Gaussian splatting, and human-aware background synthesis to produce highly varied and lifelike human sequences. Based on {\proposed}, we present SportPAL, a large-scale synthetic dataset spanning three sports— baseball, icehockey, and soccer. Together, {\proposed} and SportPAL provide a scalable foundation for constructing synthetic datasets tailored to in-the-wild human-centric vision tasks, with no need for manual 3D modeling or scene design.
\end{abstract}

\input{sec/1_intro}
\input{sec/2_related_works}
\input{sec/3_method}
\input{sec/4_dataset}
\input{sec/4_experiments}
\input{sec/5_conclusion}
\input{sec/6_acknowledgement}

\newpage

{\small
\bibliographystyle{ieee_fullname}
\bibliography{egbib}
}

\end{document}

%% file: sec/1_intro.tex
\section{Introduction}\label{sec:intro}




Data collection and annotation are critical for any computer vision task \cite{bedlam, annotatexr, annovate}. The quality and quantity of the collected data directly influence the accuracy and reliability of the trained models \cite{agora, surreal, annotatexr, qual1}. However, the process of data collection and annotation can be challenging, especially in the context of human-centric tasks, where the complexity and dynamism of human postures, coupled with variations in appearance, make it difficult to capture the full range of relevant information \cite{tokenclipose}. 

Due to the aforementioned issue, synthetic data is becoming increasingly popular for its ease of acquisition and well-annotated nature \cite{agora, surreal, bedlam, gta, hspace}. However, current synthetic data fall short in several key aspects necessary for broad applicability. For example, most existing approaches \cite{agora, surreal, pumarola20193dpeople} rely on a library of 3D assets \cite{adobe-fuse, makehuman}, pre-defined motions \cite{mixamo, amass, caeser, cmu_mocap}, or constrained rendering pipelines \cite{blender, unreal} that result in repetitive appearances, restricted visual diversity, and oversimplified motion dynamics. Additionally, manually constructing large-scale synthetic datasets with meaningful variation in body types, actions, viewpoints, clothing, and lighting remains a labor-intensive and inflexible process, making it difficult to scale to diverse domains like sports or outdoor activities. Table \ref{tab:datasets} provides a detailed summary of existing datasets (real, rendered, and composite) for human-centric vision tasks.

Given the limitations of existing datasets, we ask: \textit{Can we fully automate the synthesis of lifelike human animation from motion to scene, without any manual 3D modeling?}


Our answer is \textbf{yes}. In this paper, we introduce {\proposed}, a fully automated pipeline for synthesizing diverse, lifelike 4D human animation datasets, combining dynamic motion with photorealistic scene generation.  {\proposed} leverages Gaussian splatting and text-conditioned diffusion models to generate realistic human sequences directly from text prompts. Motion trajectories are obtained through an expert-driven motion modeling module that extracts pose data from publicly available internet videos. The framework supports scalable creation of synthetic datasets with controllable variations over body shape, clothing, motion type, camera viewpoint, and environment. 

With {\proposed}, we introduce SportPAL (Figure \ref{fig:teaser}), a large-scale synthetic dataset spanning three sports— baseball, icehockey, and soccer. SportPAL includes 2D bounding boxes, 2D and 3D pose annotations, SMPLX-based human modeling parameters, segmentation masks, and action labels. The dataset comprises over half a million frames of human animations covering a wide range of actions, camera viewpoints, lighting conditions, and subject appearances, tailored for sports scenarios. Experimentation on the SportPAL dataset demonstrates its potential for human-centric tasks. Additionally, we evaluate the cross-sport generalization ability of a pose estimator trained on our synthetic data.

Our contributions can be summarized as follows: (1) We propose {\proposed}, a fully automated pipeline for generating lifelike human models with realistic animations and backgrounds. (2) Based on {\proposed}, we introduce SportPAL, a large-scale synthetic dataset spanning three sports- baseball, icehockey, and soccer, with rich annotations specifically designed for human-centric vision tasks.



%% file: sec/2_related_works.tex
\begin{table*}[t]
    \centering
    \begin{tabular}{lccccccccc}
        \hline
        Dataset & Environment & Subjects & Keypoints & Poses & Cameras & Markerless & FPS & Frames \\
        \hline
        Human3.6M \cite{h36m} & lab & 11 & 26 & 900K & 4 & $\times$ & 50 & 3.6M \\
        MPI-INF-3DHP \cite{mpi3dhp} & lab \& outdoor & 8 & 28 & 93K & 14 & $\checkmark$ & 25/50 & 1.3M \\
        3DPW \cite{3dpw} & lab \& outdoor & 7 & 24 & 49K & 1 & $\times$ & 30 & 51K \\
        HumanEva-I \cite{humaneva} & lab & 6 & 15 & 78K & 7 & $\times$ & 60 & 280K \\
        HumanEva-II \cite{humaneva} & lab & 6 & 15 & 3K & 4 & $\times$ & 60 & 10K \\
        TotalCapture \cite{TotalCapture} & lab & 5 & 25 & 179K & 8 & $\times$ & 60 & 1.9M \\
        CMU Panoptic \cite{cmu_mocap} & lab & 8 & 18 & 1.5M & 31 & $\checkmark$ & 30 & 46.5M \\
        AIST++ \cite{AIST++} & lab & 30 & 17 & 1.1M & 9 & $\checkmark$ & 60 & 10.1M \\
        ASPset-510 \cite{ASPset} & outdoor & 17 & 17 & 110K & 3 & $\checkmark$ & 50 & 330K \\
        SportsPose \cite{SportsPose} & lab \& outdoor & 24 & 17 & 177K & 7 & $\checkmark$ & 90 & 1.5M \\
        WorldPose \cite{jiang2024worldpose} & soccer pitch & - & 24 & 2.5M & - & $\checkmark$ & 50 & 150K \\
        AthletePose3D \cite{yeung2025athletepose3d} & lab \& ice rink & 8 & 55/86 & 165K & 4/8/12 & $\checkmark$ & 60/120 & 1.3M \\
        HSPACE \cite{hspace} & rendered & 100 & GHUM \cite{ghum} & ~5M & 5 & $\checkmark$ & 60/120 & 1M \\
        GTA-Human \cite{gta} & game & $>$600 & 24 & 1.4M & - & $\checkmark$ & N/A & 1.4M \\
        3DPeople \cite{pumarola20193dpeople} & composite & 80 & 3D joints & 2.5M & 4 & $\checkmark$ & N/A & 2.5M \\
        AGORA \cite{agora} & rendered & $>$350 & 24/ 55 & 173K & 1 & $\checkmark$ & N/A & 18K \\
        BEDLAM \cite{bedlam} & rendered & 217 & 55 & - & - & $\checkmark$ & N/A & 380K \\
        \textbf{SportPAL (ours)} & synthetic & 50 & 55 & 583K & 18 & $\checkmark$ & N/A & 583K \\
        \hline
    \end{tabular}
    \caption{\textbf{Comparison of the SportPAL dataset with other datasets for human-centric vision tasks.}}
    \label{tab:datasets}
\end{table*}

\section{Related Work}\label{sec:related-work}

\subsection{Real Datasets}

Real-world images are inherently diverse, complex, and abundant. Many existing 2D pose datasets, such as COCO \cite{coco} and LSP \cite{lsp}, rely on these real images and require manual annotation of 2D poses. These datasets are often limited by the subjectivity and potential inconsistencies of human annotators, who may provide annotations that are constrained to the 2D plane and lack 3D information \cite{tokenclipose}.

While 2D pose annotation is relatively straightforward, the challenge of 3D HPE has necessitated different approaches due to the inherent ambiguity of the task \cite{mitigatingblur}. To address this, researchers have relied primarily on two approaches: marker-based and markerless motion capture systems \cite{yeung2025athletepose3d, h36m, mpi3dhp, humaneva, AIST++, ASPset}. Marker-based systems \cite{h36m, humaneva, TotalCapture} offer accurate groundtruth but are limited by the number of subjects, clothing variations, and controlled environments. Marker-less systems \cite{mpi3dhp, SportsPose, yeung2025athletepose3d, jiang2024worldpose} provide greater flexibility in terms of subject variability and real-world scenarios, but often suffer from less accurate groundtruth compared to marker-based methods. 

All existing datasets demonstrated limitations in terms of complexity, pose diversity, and the accuracy of groundtruth data \cite{h36m, mpi3dhp, humaneva}. {\proposed} addresses these challenges by leveraging text prompts and publicly available internet videos to generate a wide range of diverse and complex human-centric datasets.

\subsection{Synthetic Datasets}

Computer graphics has the potential to synthesize large-scale image datasets, where groundtruth is generated by animating human models such as SMPL \cite{smpl} or GHUM \cite{ghum}. PanopTOP \cite{panoptop} is a human-only synthetic dataset aimed at solving the distribution gap in real datasets that encompass a majority of frontal views. SURREAL \cite{surreal} was one of the first works to open-source large-scale synthetic datasets for many different human-centric tasks, including 2D/3D pose estimation, depth estimation, optical flow, and segmentation masks. Similarly, AGORA \cite{agora} proposes a synthetic dataset with high realism and highly accurate groundtruth. They aim to solve the generalizability issue by including person-person occlusions, environmental occlusions, crowds, and children's data as part of their synthetic dataset.

Although the above works were promising, they were used in addition to real datasets to get the best performance. BEDLAM \cite{bedlam} was the first work that demonstrated that neural networks trained on synthetic data can achieve state-of-the-art accuracy in 3D human pose and shape estimation. 



While these works mark promise, most of these datasets are constrained by the motion source obtained primarily from AMASS \cite{amass}, CEASER \cite{caeser}, or CMU MoCap \cite{cmu_mocap} datasets. Similarly, they tend to use a database of skin tones, hair, motion sequences, and clothing modeled using software like Blender \cite{blender} or Unreal Engine \cite{unreal}. This limits their \textit{applicability to in-the-wild dynamic scenarios}, which often require complex motion with varying body types, clothing, and environments. Thus, in contrast, our work {\proposed} is a fully automated text-guided 4D human animation system, through which we create SportPAL — a large-scale, diverse, and realistic synthetic dataset tailored for in-the-wild human activities.





%% file: sec/3_method.tex
\begin{figure*}[t]
  \centering
  \includegraphics[width=\linewidth]{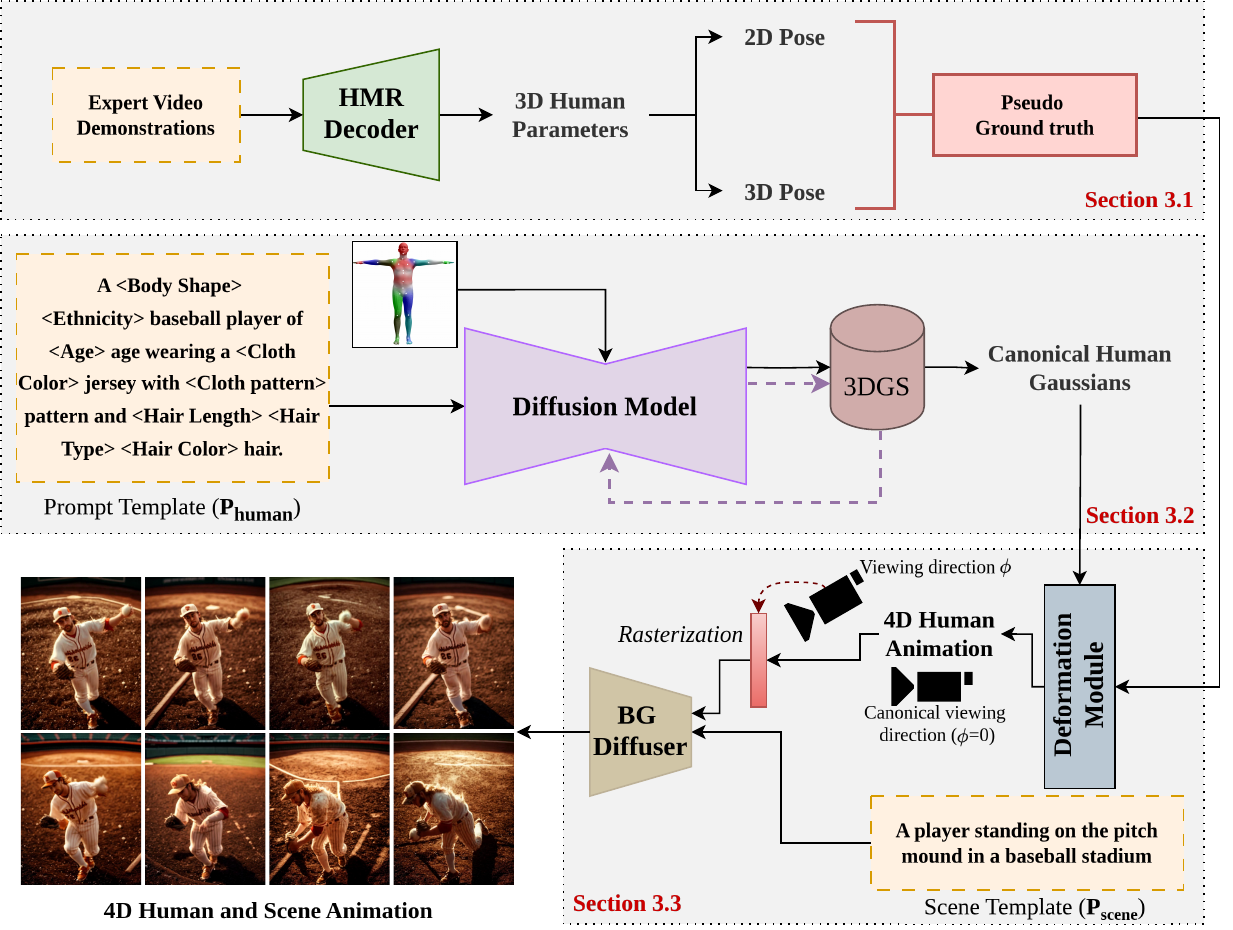}
  \vspace{-20px}
   \caption{\textbf{The pipeline of {\proposed}.} Our method consists of three stages: \textcolor{Red}{\textbf{Section 3.1}} \textbf{Motion Extraction} extracts 3D human motion from expert demonstrations; \textcolor{Red}{\textbf{Section 3.2}} \textbf{Canonical Human Gaussians} generates diverse human avatars via diffusion-guided Gaussian optimization; and \textcolor{Red}{\textbf{Section 3.3}} \textbf{Scene Composition} deforms avatars with motion, rasterizes them from varied viewpoints, and synthesizes realistic backgrounds using a scene-conditioned diffuser.}
   \label{fig:overview}
\end{figure*}

\section{Method} \label{sec:method}

In this section, we describe our end-to-end pipeline for generating motion-driven, photorealistic human avatars in diverse environments. Our method begins by extracting 3D motion data from real video demonstrations, capturing realistic body dynamics. We then generate a wide range of human avatars with varying appearances using prompt-driven diffusion-guided optimization. These avatars are then animated with the extracted motion, rendered from multiple camera viewpoints, and composited into realistic backgrounds that match the human pose context. This results in high-quality, fully synthetic human sequences suitable for training and evaluating vision models. An overview of the pipeline is shown in Figure~\ref{fig:overview}.

\subsection{Motion Extraction} \label{sec:motion_rep}

Given a set of expert video demonstrations from publicly available internet videos, we utilize a pretrained human mesh recovery method \cite{smplerx} to extract the SMPLX \cite{smpl} pose and shape parameters of the expert. Since each demonstration features a single focal expert, we normalize the shape parameters across the sequence. Following prior works \cite{pose2mesh, d2ahmr, hmr}, we regress these parameters to 3D joint coordinates using a predefined regression matrix $G \in \mathbb{R}^{K \times M}$, where $K$ and $M$ represent the number of joints and SMPL mesh vertices, respectively. The resulting 3D joints are then projected onto the image plane to obtain the 2D joint coordinates. We specifically opt for expert demonstrations to encode motion representations, rather than relying on language-driven models \cite{mdm, flowmdm, priormdm}, as these models often fail to generate diverse and realistic motion sequences for complex, in-the-wild actions \cite{sahili2025textdrivenmotiongenerationoverview}.

An important characteristic to note is that the accuracy of the extracted motion from the mesh recovery method \cite{smplerx} doesn't need to perfectly match the observed pose of the person in the original video. Since our pipeline uses the estimated pose as the pseudo-motion to drive synthetic avatar animation (i.e., independent of the original viewpoint), we will still be able to generate perfect image–pose pairs in entirely new camera configurations for the corresponding rendered synthetic image. 

\begin{figure*}[t]
  \centering
  \includegraphics[width=\linewidth]{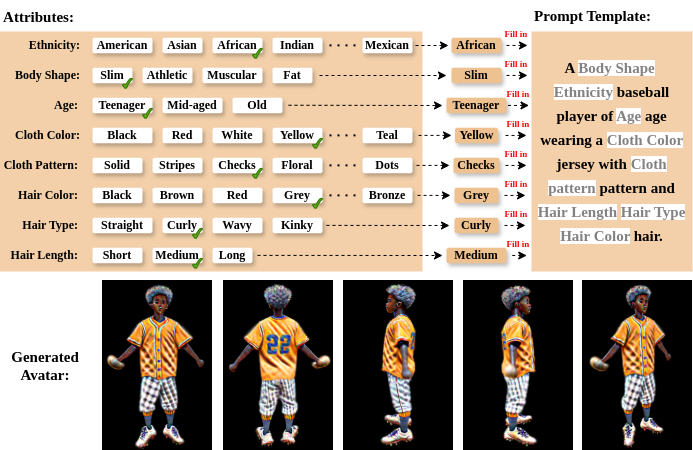}
  \caption{\textbf{Prompt modeling pipeline of {\proposed}.} Diverse prompt templates are used to synthesize a variety of 3D canonical avatars, illustrated here from five different camera views.}
  \label{fig:prompt}
\end{figure*}

\subsection{Canonical Human Gaussians} \label{sec:canonical_gaussian}

Most prior works \cite{gauhuman, 3dgs-avatar, gaussianavatar} use images or 3D scans to guide the initialized Gaussians towards realistic human representations. This limits the diversification of the model to the available visual data given as input. Thus, in this work, we optimize the Gaussian parameters using the synthesized RGB ($x_c$) and depth map ($d_c$) from a pretrained diffusion backbone \cite{humangaussian} which is guided by text prompts. This approach is specifically chosen to introduce variability and flexibility in the generated human avatars. 

Specifically, we first initialize a SMPLX human model in canonical space. Then we initialize the Gaussians on the canonical representation instead of initializing with random points or structure-from-motion (SfM) \cite{sfm}. This ensures dense point distributions for initialization. Then, we utilize a pretrained diffusion model that is trained to simultaneously denoise RGB ($\mathbf{x_{c}}$) and depth map ($\mathbf{d_{c}}$) conditioned on a canonical 3D pose and prompt template ($\textbf{P}_{\text{human}}$). 

Based on this, a dual-branch Score Distillation Sampling (SDS) \cite{humangaussian} is used to optimize the human appearance and geometry of the initialized canonical representation. The loss can be computed as shown in Equation \eqref{eq:sds}.

\begin{equation}
    \begin{aligned}
\nabla_{\theta} \mathcal{L}_{\mathrm{SDS}} & =\lambda_{1} \cdot \mathbb{E}_{\boldsymbol{\epsilon}_{\mathbf{x_{c}}}, t}\left[w_{t}\left(\boldsymbol{\epsilon}_{\phi}\left(\mathbf{x}_{c} ; {p_{c}}, \textbf{P}_{\text{human}}\right)-\boldsymbol{\epsilon}_{\mathbf{x_{c}}}\right) \frac{\partial \mathbf{x_{c}}}{\partial \theta}\right] \\
& +\lambda_{2} \cdot \mathbb{E}_{\boldsymbol{\epsilon}_{\mathbf{d_{c}}}, t}\left[w_{t}\left(\boldsymbol{\epsilon}_{\phi}\left(\mathbf{d}_{c} ; {p_{c}}, \textbf{P}_{\text{human}}\right)-\boldsymbol{\epsilon}_{\mathbf{d_{c}}}\right) \frac{\partial \mathbf{d_{c}}}{\partial \theta}\right]
    \end{aligned}
    \label{eq:sds}
\end{equation}

\noindent where $\lambda_{1}$ and $\lambda_{2}$ are coefficients that balance the effects between RGB and depth. We compute $\mathcal{L}_{\mathrm{SDS}}$ according to our parsed random render $\mathcal{R}^{'}$ for supervision. $\epsilon_\phi(\cdot)$ are the $\epsilon$-predictions derived from the joint texture-structure model, $p_{c}$ correspond to the canonical pose map and $w_t$ is the noise sampler.

\paragraph{Prompt Modeling ($\textbf{P}_{\text{human}}$).} To systematically generate diverse descriptions (template) for synthetic data generation, we designed a prompt modeling framework. This framework incorporates a comprehensive set of attributes, including ethnicity, hair type, hair length, hair color, clothing color, clothing patterns, body types, and age groups, to generate realistic and varied character templates as shown in Figure \ref{fig:prompt}. Each attribute was treated as a parameter within a cyclic iterator, ensuring balanced sampling and avoiding repetitive combinations. The framework dynamically combines these attributes into descriptive sentences (template) tailored to specific scenarios based on the activity. 

\begin{figure*}[t]
  \centering
  \includegraphics[width=\linewidth]{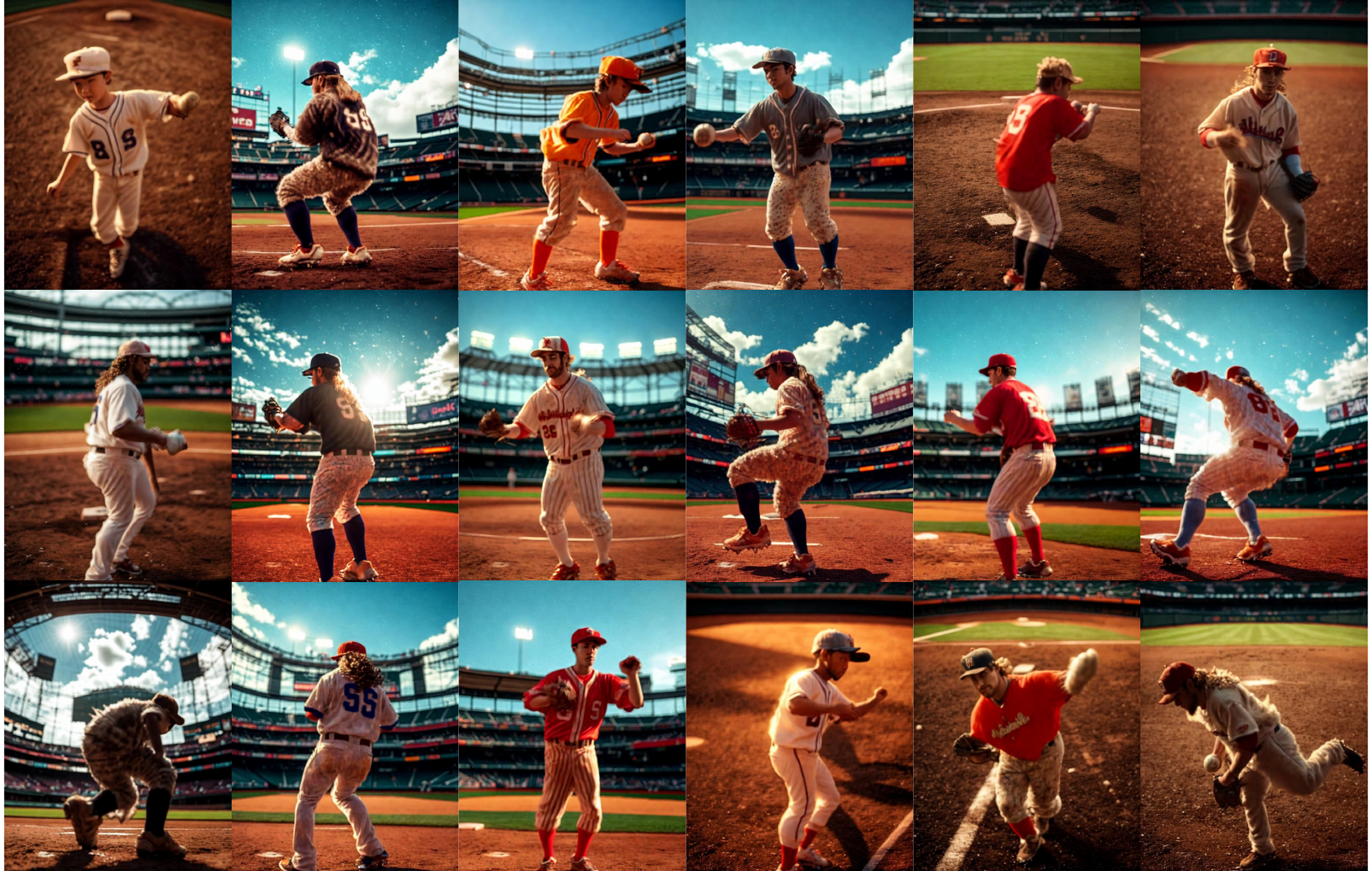}
  \caption{\textbf{Qualitative examples of synthetic images.} Using {\proposed}, we applied diverse prompt modeling to generate images across the three sports categories included in the SportPAL dataset.}
   \label{fig:qual}
\end{figure*}

\subsection{Scene Composition} \label{sec:Backgrounddiffuser}

This subsection is organized into the following stages: (1) The canonical human Gaussian is deformed using expert motion sequences; (2) The deformed Gaussians are rasterized from a specified camera viewing direction ($\phi$); and (3) A background diffusion algorithm generates the corresponding background for the rasterized Gaussian, guided by a scene prompt ($\textbf{P}_\textbf{scene}$).

\noindent \textbf{Deformation Module.} The generated canonical avatars from Section \ref{sec:canonical_gaussian} are deformed based on the motion sequences computed from Section \ref{sec:motion_rep}. Given an input motion sequence, each frame provides body pose parameters and a global orientation. These parameters are used to update the mesh vertices to represent the current human pose. Following literature \cite{humangaussian, gauhuman} which utilizes the precomputed face associations, barycentric coordinates, and local surface normals, the positions of the Gaussians are recomputed at each frame, ensuring that the deformation of the Gaussian cloud closely follows the articulated motion of the underlying mesh. To maintain fidelity during deformation, points with excessive deviation between the expected and actual projected positions are culled. This filtering ensures stability and avoids visual artifacts due to misalignments during rapid or complex motions.

\noindent \textbf{Rasterization.} Once the Gaussians have been deformed according to the target pose, they are rasterized into a 2D image plane. Following standard 3D Gaussian splatting methods \cite{3dgs}, each Gaussian is projected using perspective projection based on the camera's intrinsic and extrinsic parameters. The final image at each pixel is computed by blending Gaussian contributions weighted by opacity and spatial extent, enabling photorealistic rendering without aliasing.

\noindent \textbf{Camera Control.} The camera view is dynamically adjusted to capture the deformed model from different viewpoints. Instead of continuously orbiting around the model, the camera position and orientation (elevation and azimuth parameters) are randomly updated for each motion sequence to achieve the desired perspective. The intrinsic parameters, such as field of view, aspect ratio, and near/far clipping planes, remain fixed. 

\noindent \textbf{Background Diffusion.} The diffusion module accommodates two tasks: removing artifacts from the rasterized avatars and inducing relevant background in the rasterized human projection. This process results in realistic avatars with scenes as depicted in Figure \ref{fig:qual}, where the output shows scenes augmented with the rasterized humans. 

We utilize IC-Light \cite{iclight} as the backbone for background generation through its physically grounded light transport consistency principle. This approach ensures that intrinsic object properties, such as albedo and reflectance colors, remain unaltered while accurately modifying illumination effects. This blending of the rasterized avatars with dynamically generated backgrounds results in realistic generated data. The algorithm is driven by an objective function denoted as $\mathcal{L}_\textrm{diff}$ as defined in Equation \eqref{eq:iclight_loss_all}.

\begin{align}
    \mathcal{L}_\textrm{diff} = & \; \lambda_{v} \Big[\| \varepsilon - \delta (\epsilon(I_L)_t, L, t, \epsilon(I_d)) \|^2_2 \Big] \nonumber \\
    & + \lambda_{ic} \Big[\| M \odot (\epsilon_{L_1+L_2} - \delta (\epsilon(L_1, L_2))) \|_2^2 \Big]
    \label{eq:iclight_loss_all}
\end{align}

\begin{table*}
\caption{\small {\textbf{Performance of 2D HPE techniques across the SportPAL dataset.} The best result is highlighted in \textbf{bold} format.}}
\centering
\adjustbox{width=0.75\linewidth}{
\setlength{\tabcolsep}{5pt}
\begin{tabular}{lccccccccc}
\toprule
\multirow{2}{*}{Method} 
& \multicolumn{3}{c}{Baseball} 
& \multicolumn{3}{c}{Icehockey} 
& \multicolumn{3}{c}{Soccer} \\
\cmidrule(lr){2-4}
\cmidrule(lr){5-7}
\cmidrule(lr){8-10}
& AP$^5\textcolor{red}{\uparrow}$ & AP$^{10}\textcolor{red}{\uparrow}$ & AP$^{15}\textcolor{red}{\uparrow}$ 
& AP$^5\textcolor{red}{\uparrow}$ & AP$^{10}\textcolor{red}{\uparrow}$ & AP$^{15}\textcolor{red}{\uparrow}$ 
& AP$^5\textcolor{red}{\uparrow}$ & AP$^{10}\textcolor{red}{\uparrow}$ & AP$^{15}\textcolor{red}{\uparrow}$ 
\\ 
\midrule
DETR-based \cite{detr}                 & 64.33 & 74.61 & 77.19 & 45.01 & 61.7 & 70.46 & 64.85 & 81.04 & 88.20 \\
TokenPose~\cite{tokenhmr} & \textbf{80.74} & \textbf{89.35} & \textbf{93.68} & \textbf{62.75} & \textbf{96.92} & \textbf{99.91} & \textbf{67.28} & \textbf{92.56} & \textbf{98.51} \\
\bottomrule
\end{tabular}
}
\label{tab:exp:main}
\end{table*}

\noindent where $\lambda_v$ and $\lambda_{ic}$ are the weights assigned to 1 and 0.1 by default, respectively. $M$ denotes the foreground mask, $L_1$ and $L_2$ are the illumination masks, $I_L$ denotes the appearance image, and $L$ represents the environment illumination. $\varepsilon$ is the latent representation of an image, $\delta$ is the network used to predict noise, $\epsilon$ is the diffusion target and $I_d$ represents the degradation image.

The technique involves imposing constraints based on the linearity of light transport, such that the illumination of objects under mixed lighting is consistent with their lighting conditions \cite{debevec2000acquiring}. This constraint is integrated into the diffusion-based generator, allowing for precise control over foreground and background lighting, including shadow induction for enhanced realism. 

%% file: sec/4_dataset.tex
\section{Dataset} \label{sec:datasets}

Based on {\proposed}, we present the largest dataset featuring the most diverse range of actions from various sports, termed as SportPAL, short for Sport Pose and Action Library. This dataset includes activities from three different sports, including baseball, icehockey, and soccer. It provides 2D bounding boxes, groundtruth 2D and 3D human pose annotations, SMPLX human pose and shape parameters, along with the action class. Comprising more than 2,000 videos with 50 subjects representing a wide range of body types and ethnicities, and over 500,000 frames, this dataset offers a rich resource for research on various in-the-wild human-centric vision tasks. Table \ref{tab:sportspal-data-split} shows the data split of SportPAL categorized based on the sport.

\begin{table}[t]
\centering
\caption{\small \textbf{SportPAL Data Split.} Data statistics across three sports collected using the proposed {\proposed} framework. 'frames-aug' denotes total frames after augmentations. }
\adjustbox{width=0.9\linewidth}{
\setlength{\tabcolsep}{6pt}
\begin{tabular}{lcccc}
\toprule
\textbf{Sport} & \textbf{Split} & \#\textbf{Subjects} & \#\textbf{Clips} & \#\textbf{Frames} \\ 
\midrule
\multirow{4}{*}{Baseball} 
    & Train  & 15 & 1,000 & 253,869 \\
    & Valid  & 15 & 304   & 80,810  \\
    & Test   & 5  & 300   & 71,875  \\
\midrule
\multirow{4}{*}{Icehockey} 
    & Train  & 10 & 195 & 75,468 \\
    & Valid  & 10 & 50 & 18,867 \\
    & Test   & 5 & 12 & 7,487 \\
\midrule
\multirow{4}{*}{Soccer} 
    & Train  & 10 & 116 & 57110 \\
    & Valid  & 10 & 30 & 14,277 \\
    & Test   & 5 & 5 & 3639 \\
\midrule
\textbf{Total} & - & \textbf{50} & \textbf{2,012} & \textbf{583,403} \\ 
\bottomrule
\end{tabular}
}
\label{tab:sportspal-data-split}
\end{table}

%% file: sec/4_experiments.tex
\section{Experiments}\label{sec:exp}

\subsection{Training Details}

The Gaussian model is initialized with 100k samples on the SMPLX mesh with an opacity set to 1. Colors are represented using Spherical Harmonics (SH) coefficients of degree 0. The bidirectional SDS loss weights $\lambda_{1}$ and $\lambda_{2}$ are both set to be 0.5. The training resolution is set to be 1024 with a batch size of 8. The training of the canonical Gaussian is done for 3600 iterations, and it takes about two hours on a single GPU of NVIDIA A6000 (48GB) GPU. The adaptive densifying and pruning is done from 300 to 800 iterations at an interval of 100 steps. 

\paragraph{Augmentations.} We vary the virtual camera parameters for each sequence to capture a wide range of viewing perspectives to improve generalization for arbitrary camera orientations. Specifically, we adjust the azimuth angle (i.e., rotating the camera horizontally around the subject) to simulate different viewpoints.

\begin{table}[t]
\caption{\small \textbf{Quantitative comparison of the impact of finetuning on the Icehockey and Soccer datasets.} The best result is highlighted in \textbf{bold} format.}
\centering
\adjustbox{width=0.85\linewidth}{
\setlength{\tabcolsep}{5pt}
\begin{tabular}{lcccc}
\toprule
\textbf{Sport} & \textbf{Method} & AP$^5\textcolor{red}{\uparrow}$ & AP$^{10}\textcolor{red}{\uparrow}$ & AP$^{15}\textcolor{red}{\uparrow}$ \\
\midrule
\multirow{3}{*}{Icehockey} 
    & w/o finetuning  & 62.75 & 96.92 & 99.91 \\
    & w/ finetuning   & \textbf{63.47} & \textbf{98.10} & \textbf{99.98} \\
    &                 & \textcolor{red}{(+0.72)} & \textcolor{red}{(+1.18)} & \textcolor{red}{(+0.07)} \\
\midrule
\multirow{3}{*}{Soccer} 
    & w/o finetuning  & 67.28 & 92.56 & 98.51 \\
    & w/ finetuning   & \textbf{71.46} & \textbf{94.68} & \textbf{99.13} \\
    &                 & \textcolor{red}{(+4.18)} & \textcolor{red}{(+2.12)} & \textcolor{red}{(+0.62)} \\
\bottomrule
\end{tabular}
}
\label{tab:finetune_comparison}
\end{table}

\begin{figure*}[t]
  \centering
  \includegraphics[width=1\linewidth]{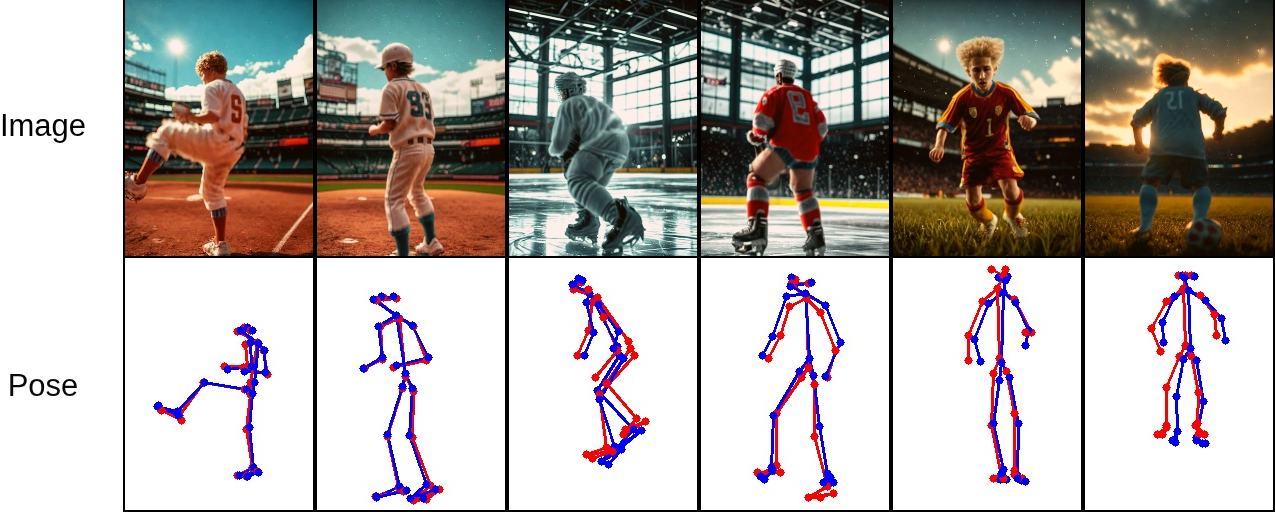}
   \caption{\textbf{Qualitative visualizations of pose estimation results generated by TokenPose \cite{tokenpose}.} Groundtruth and predicted keypoints are shown in \textcolor{red}{\textbf{red}} and \textcolor{blue}{\textbf{blue}}, respectively.}
   \label{fig:vis}
\end{figure*}

\subsection{Preliminary Results}


In this section, we present two pivotal experiments designed to demonstrate the quality of the SportPAL dataset. Specifically, we evaluate the performance of 2D HPE models on the SportPAL dataset—both for training and testing—across the three sports: baseball, ice hockey, and soccer. We assess the efficacy of two different pose estimation architectures and explore the impact of fine-tuning across different sport domains.

\paragraph{Pose Estimation Performance on SportPAL.} Table \ref{tab:exp:main} summarizes the quantitative results of 2D HPE on the SportPAL dataset for each of the three sports. We trained and evaluated two models: (1) A baseline pose estimator based on a transformer architecture derived from Detection Transformers (DETR) \cite{detr}; and (2) the TokenPose model \cite{tokenpose}, a transformer-based pose estimator designed for dense spatial correspondence learning.

Both models were trained individually on each sport subset of the SportPAL dataset using the same experimental settings for consistency and comparability. As shown in Table \ref{tab:exp:main}, TokenPose consistently outperforms the baseline model across all sports, demonstrating superior spatial reasoning and robustness to sport-specific variations in appearance, motion, and occlusion. The improvements are particularly notable in icehockey where the baseline model failed considerably. Some visualization results of TokenPose on test set of the respective sports category are shown in Figure \ref{fig:vis}. 

\paragraph{Cross-Sport Fine-Tuning.} To further demonstrate the benefit of data diversity in SportPAL, we conducted a cross-sport fine-tuning experiment, as shown in Table \ref{tab:finetune_comparison}. We initially trained the TokenPose model on the baseball subset, which is the largest among the three categories. Subsequently, we fine-tuned this pretrained model separately on the icehockey and soccer subsets, which have comparatively less training data. Compared with directly training on icehockey or soccer data, consistent improvement across both sports was observed after fine-tuning. This result shows that the diversity of our generated data can be benefit for model's generalizability. 

However, the degree of improvement varied. Icehockey showed marginal gains, likely due to the distribution gap between baseball and icehockey, while notable improvements were observed in soccer. This could be due to the nature of sports, where baseball and soccer involve running-based locomotion, whereas icehockey features skating, which results in significantly different body posture and joint articulation. Additionally, icehockey players usually have bulky gear, occluding the player's joint positions. Thus, this domain disparity limits the effectiveness of transferring learned representation from baseball to icehockey. 

\noindent \textbf{Qualitative Results.} Figure \ref{fig:qual} illustrates some synthetic images from SportPAL dataset generated using {\proposed} framework. Some notable aspects include the shadows of the human reflected on the scene, making it more realistic. The illustrations also show varying lighting conditions, camera angles, and varying cloth appearances.  


%% file: sec/5_conclusion.tex
\section{Conclusion}\label{sec:discussion}

In this work, we introduced {\proposed}, a fully automated and scalable pipeline for generating lifelike 4D human animations with rich diversity across motion, appearance, and environmental conditions, without the need for manual 3D modeling or scene design. Leveraging this framework, we created SportPAL, the largest synthetic dataset for sports-focused, human-centric vision tasks, complete with high-quality annotations across multiple modalities. Our experiments validate the effectiveness of both the dataset and the underlying generation pipeline, demonstrating potential for advancing HPE and related tasks in challenging real-world scenarios where traditional data collection is limited.

\paragraph{Future Work.} While {\proposed} offers a flexible framework for generating diverse and realistic 4D human animations, several promising directions remain for future exploration. First, enforcing temporal consistency across background generation could enable the creation of dynamic, coherent scenes, further supporting downstream video-based applications \cite{zhang2024monst3r, chen2025easi3r}. Second, optimizing the pseudo groundtruth extracted from expert demonstrations with physics-based constraints could reduce motion jitter and improve the physical plausibility of pose predictions, enhancing the overall realism of generated sequences.

%% file: sec/6_acknowledgement.tex
\paragraph{Acknowledgement}

We sincerely thank the Baltimore Orioles of Major League Baseball for their generous support through the Mitacs Accelerate Program, which was instrumental in advancing this research. 
